\documentclass[12pt]{iopart} 

\usepackage{epsfig}

\newcommand{\beq}{\begin{equation}}
\newcommand{\eeq}{\end{equation}}
\newcommand{\beqa}{\begin{eqnarray}}
\newcommand{\eeqa}{\end{eqnarray}}

\def\la{\mathrel{\mathpalette\fun <}}

\def\fun#1#2{\lower3.6pt\vbox{\baselineskip0pt\lineskip.9pt
  \ialign{$\mathsurround=0pt#1\hfil##\hfil$\crcr#2\crcr\sim\crcr}}}

\begin{document} 

\title{Theory Challenges of the Accelerating Universe} 
\author{Eric V.\ Linder} 
\address{Berkeley Lab, University of California, Berkeley, CA 94720 USA} 
\ead{evlinder@lbl.gov}

\date{\today} 

\begin{abstract} 
The accelerating expansion of the universe presents an exciting, 
fundamental challenge to the standard models of particle physics 
and cosmology.  
I highlight some of the outstanding challenges in both developing 
theoretical models and interpreting without bias the observational 
results from precision cosmology experiments in the next decade 
that will return data to help reveal the nature of the new physics. 
Examples given focus on distinguishing a new component of 
energy from a new law of gravity, and the effect of early 
dark energy on baryon acoustic oscillations. 

\end{abstract} 


\section{Introduction} \label{sec:intro} 

Acceleration of the expansion of the universe is an implicit 
possibility within Einstein's equations of motion, achievable through 
a component of the energy density with sufficiently negative pressure. 
One example is the cosmological constant, which has close ties to 
quantum field theory and the nature of the spacetime vacuum. 
Acceleration also occurs during the epoch of inflation in the first 
tiny fraction of the age of the universe, but here is believed to 
arise from a dynamical field connected with high energy physics. 

With the discovery of the recent acceleration of the cosmic expansion 
a fundamental question is whether we are seeing physics from a cosmological 
constant (whose magnitude we are far from understanding) or a new, 
dynamical field -- or completely novel physics such as an expanded 
theory of gravity.  We can hope that by studying this modern acceleration 
epoch we can probe uncharted areas of physics, learning about 
quantum and gravitational physics, and perhaps even approach a unification 
of the two. 

In this article I consider some broad theoretical avenues for 
understanding the accelerating universe and how to interpret the 
data in the next decade to reveal the nature of the new physics in 
a clean and robust manner.  In \S\ref{sec:ideas} I discuss some 
general, model independent conclusions we might draw about the dark 
physics.  \S\ref{sec:real} examines the prospects for distinguishing 
between a new physical component of energy density in the universe 
and a new physical law for gravitation, and the necessity of making 
such a distinction.  The search for cosmological methods of probing 
cleanly the nature of the acceleration is investigated from a theory 
perspective in \S\ref{sec:ivsr}, cautioning against implicitly assuming 
an answer that we are trying to find.

\section{Which Ideas?} \label{sec:ideas} 

A huge variety of theoretical ideas exist that attempt to explain the 
acceleration of our universe.  For one recent survey, see \cite{copeland}, 
as well as these proceedings.  While some of the initial ideas have 
fallen out of favor under the pressure of observations, to a large extent 
the current data sets are insufficient to decide on a particular physical 
origin.  This has led to ambitious and detailed plans for future 
experiments to guide us in our exploration. 

Here I attempt only to make some general points that might aid our 
understanding.  First, on the surface at least, dark energy (modern 
acceleration) 
and inflation (early universe acceleration) appear very different and 
we should not expect perhaps to apply the same methods of analysis to each. 
I will be bold enough to posit that dark energy may be a harder problem 
to solve than inflation.  Unlike inflation, dark energy is likely not a slow 
roll phenomenon (or not exclusively slow roll) and it does not currently 
completely dominate the universe. 

In fact, dark energy faces us with a Goldilocks problem, after the folk 
tale where from a landscape of possibilities the character Goldilocks 
found some components (e.g.\ porridge, in the tale) were too hot, some were 
too cold, and one was just right.  Dark energy is neither dynamically 
fully dominant nor negligible, but at an intermediate stage ``just right'' 
-- the dimensionless energy density $\Omega_{\rm DE}\sim 0.7$ -- to allow 
dark energy and matter cosmological observations (whereas a 
factor of four ago in expansion factor the dark energy was undetectable 
and a factor of four in the future there will be relatively few large matter 
structures within the visible universe).  

Similarly, the expansion is accelerating -- dark energy equation of state 
$w\la -0.8$ -- but one needs either a period of fast roll in the field 
or else fine tuning of initial conditions.  Inflation, by contrast, is 
simpler to treat because during inflation the field's energy density 
is completely dominant and the field is slowly rolling.  In this sense, 
measurements of inflation lend themselves to more straightforward 
interpretation in terms of the underlying physics.  

One question I will not address is the bedrock issue of the cosmological 
constant $\Lambda$; while observations can tell us whether the physics 
is distinct from the cosmological constant, they cannot declare definitely 
that the answer is the cosmological constant if the physics looks nearly 
like the 
cosmological constant.  Moreover, if the physics is not that of the 
cosmological constant we still must figure out what happened to the 
cosmological constant -- why is it zero? 

\begin{figure}[!hbt]
\begin{center} 
\psfig{file=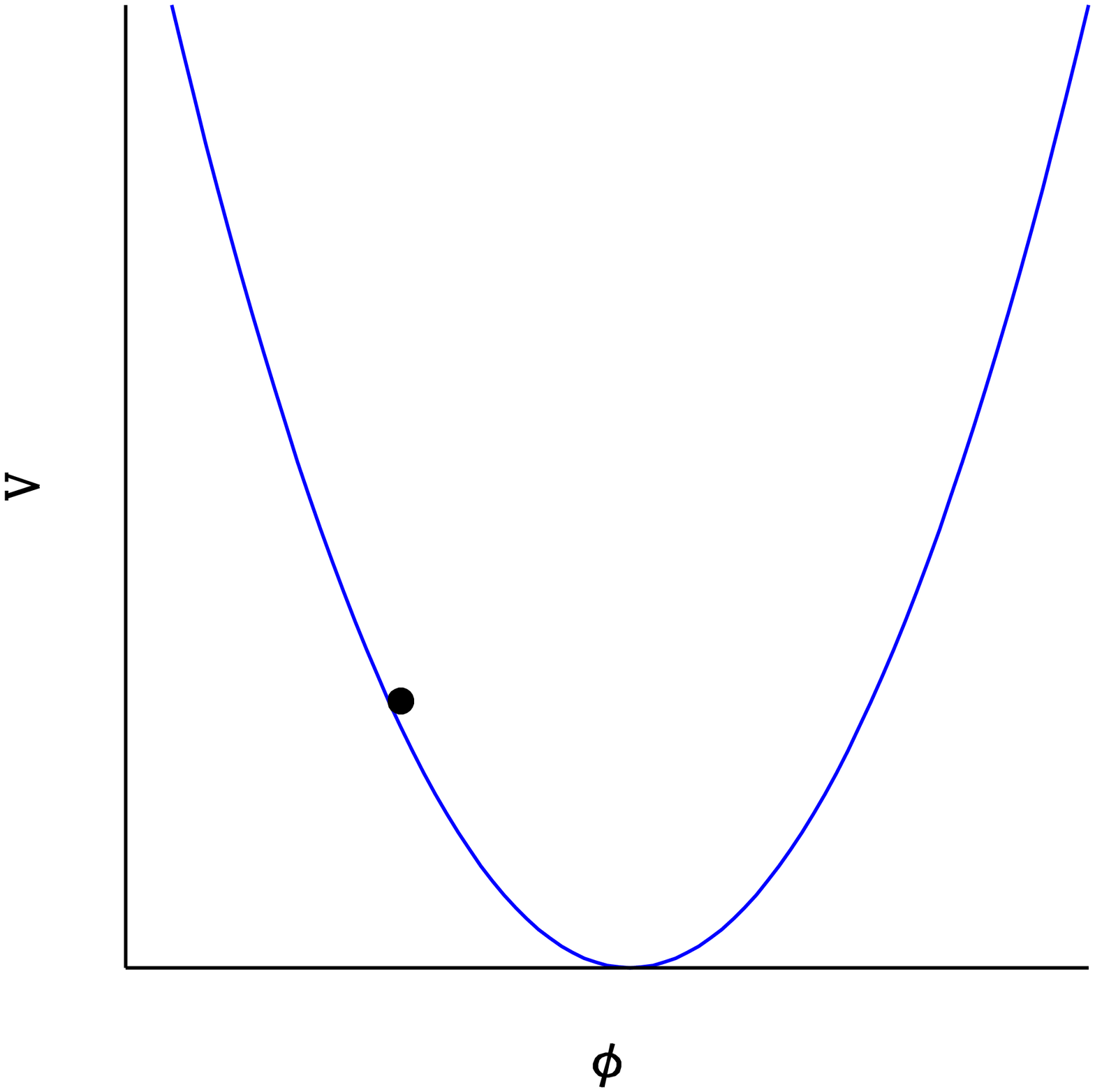,width=2.in} \qquad
\psfig{file=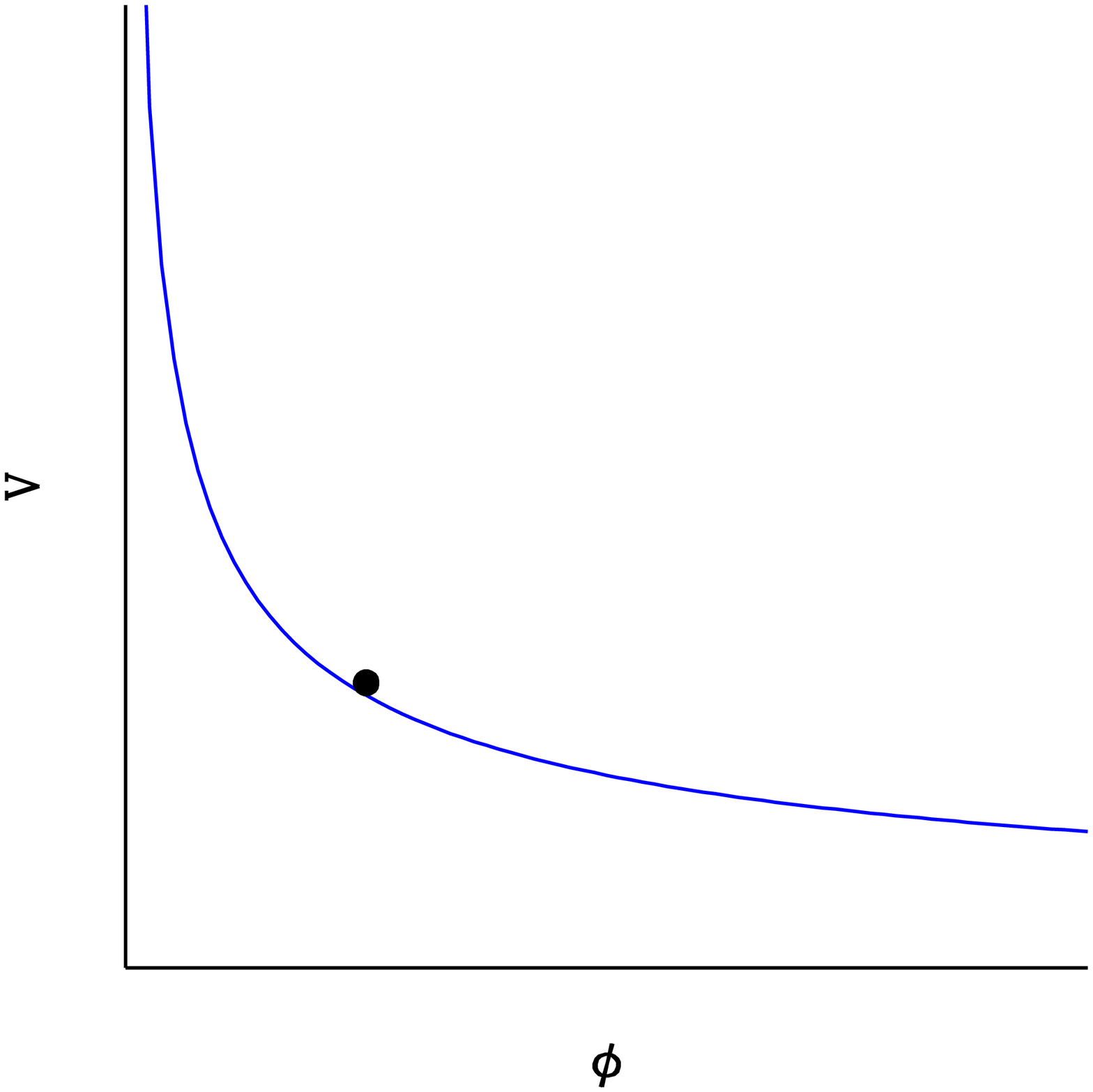,width=2.in} 
\caption{Cartoons of thawing and freezing potentials with the scalar 
field rolling down.  The left panel represents a thawing potential 
where the Hubble drag dominates early but diminishes with age so the 
field is released to roll. The right panel represents a freezing potential 
where the steepness drives the early evolution but at late times the 
Hubble drag dominates.  
}
\label{fig:pots} 
\end{center} 
\end{figure}

Beyond $\Lambda$, perhaps the simplest physics is that of a canonical 
scalar field.  While a wide variety of different potentials 
have been suggested for such a field, there are common elements and 
physics among them.  From the Klein-Gordon equation of motion, 
\beq 
\ddot\phi=-3H\dot\phi-dV/d\phi, 
\eeq 
we see that the field evolution is driven by the steepness of the 
potential $V(\phi)$ and dragged by the Hubble expansion $H=\dot a/a$. 
By the time the dark energy causes acceleration of the universe, one or 
the other of these terms will dominate.  Either the field starts 
frozen by the early, high Hubble drag and then is released to roll as $H$ 
decreases in the expansion -- we say such fields are ``leaving $\Lambda$'' 
or ``thawing'' -- or the field starts by rolling down a steep potential 
but slows to a crawl as it comes into a flatter section of the potential 
-- such fields are ``approaching $\Lambda$'' or ``freezing''.  
See Figure~\ref{fig:pots} for illustration of the two cases.

\begin{figure}[!hbt]
\begin{center} 
\psfig{file=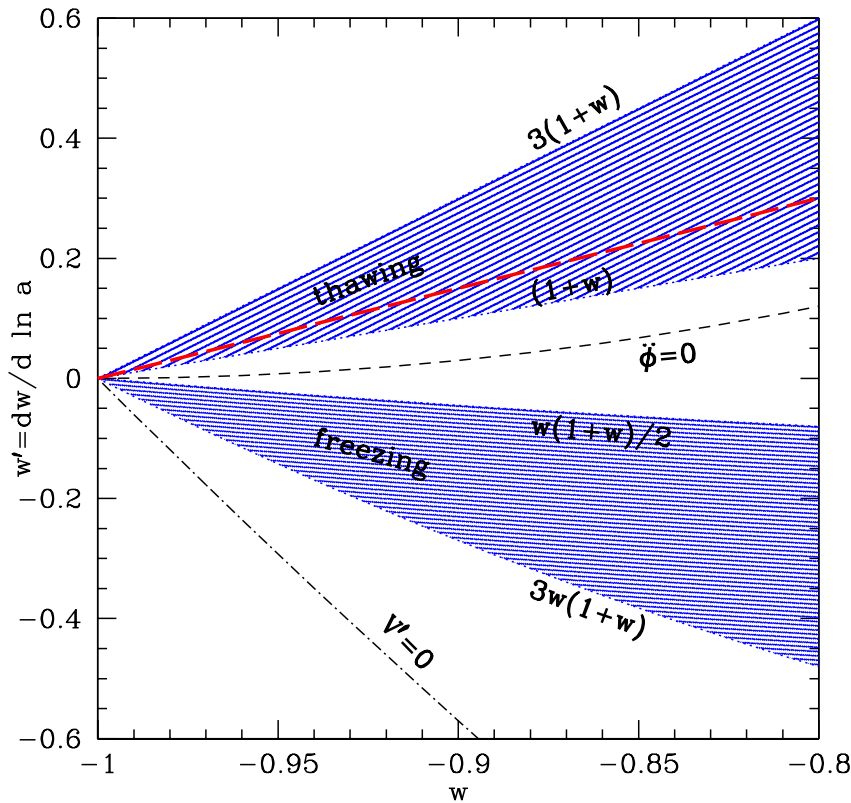,width=4.5in} 
\caption{The phase space $w'$-$w$ possesses distinct regions corresponding 
to thawing and freezing behavior.  In between, the field evolution would 
need to be finely tuned to coast, $\ddot\phi\approx0$, neither accelerating 
nor decelerating.  The long-dashed (orange) line shows the degeneracy 
surface of observations that can measure only the averaged or assumed 
constant equation of state.  (Adapted from \cite{caldlin}.) 
}
\label{fig:phase} 
\end{center} 
\end{figure}

This bimodal classification of canonical scalar field behavior 
\cite{caldlin} has proved quite successful.  The phase space behavior 
of such fields giving rise to a current epoch of acceleration falls 
into one of two distinct regions, separated by a distance $\sigma(w') 
\approx 2(1+w)$, where $1+w$ is the ``tilt'' of the equation of state 
away from the cosmological constant value of $w=-1$, and $w'=dw/d\ln a$ 
is the running.  As we see from Figure~\ref{fig:phase}, the reason 
there is a separation between the thawing and freezing behaviors is 
that for a field to live in the phase space in between, it must be 
fine tuned so the driving and dragging terms balance nearly exactly 
to yield $\ddot\phi=0$ -- over many dynamical Hubble times.

Another lesson from Figure~\ref{fig:phase} is that cosmological 
observations, which possess inherent degeneracies between parameters, 
do not zero in on a specific point in the phase space but rather an 
elongated region following the degeneracy.  The long-dashed line shows 
the degeneracy direction (virtually all types of observations have roughly 
the same degeneracy direction in the $w'$-$w$ phase space; see, e.g., 
\cite{coohutbau}).  In particular, any experiment that lacks the 
redshift reach and accuracy to be sensitive specifically to the 
time variation (such as all current and near term experiments), seeing 
only an averaged equation of state or constant $w$, will not be able to 
distinguish points along or near the long-dashed line.  

This implies that 
a result consistent with the cosmological constant is also consistent 
with almost the entire half of the populated phase space that 
represents the thawing models.  So near term experiments that might 
deliver a measurement of dark energy in terms of constant $w$ to 5\%, 
say $\langle w\rangle=-1\pm0.05$, would not in fact lead us to the 
conclusion that the answer is the cosmological constant.  A 
new generation of experiments specifically designed to be sensitive 
to time variation $w'$ is required to gain insight into the physics 
of the acceleration: cosmological constant or not, thawing or freezing. 

The characteristics of approaching or leaving $\Lambda$, and the 
cosmological physics defining boundaries for the associated freezing 
or thawing behaviors, are general enough that they apply to many other 
explanations of acceleration besides canonical scalar fields \cite{paths}. 
An effective equation of state can be defined for any Friedmann 
expansion equation, even if the modification arises from alteration 
of the gravity theory rather than a physical component.  An example 
of this is the braneworld theory \cite{dgp,ddg}, which follows a 
freezing track in the effective phase space.  It is awe inspiring 
to think that next generation cosmological measurements can put 
precision limits on something as exotic and fundamental as the 
five dimensional Planck mass, yet the Supernova/Acceleration Probe 
(SNAP: \cite{snap}) could determine $M_5/(H_0 M_{\rm Pl}^2)^{1/3}$ to 
0.2\%.

\section{Which Reality?} \label{sec:real} 

While the expansion history, or equivalently the effective dark energy 
equation of state phase space, description can treat a physical component 
such as a scalar field or a physics modification such as an extended theory of 
gravity equally, we of course also want to know which is the true 
physical origin of the cosmic acceleration.  This requires experimental 
data giving both the expansion history and the growth history 
of mass density fluctuations.  

In general relativity, the two histories are 
tied together; one can write the growth purely in terms of the expansion. 
For other theories of gravity, however, the growth behavior is governed 
both by the expansion and by the specifics of the gravity theory.  
Therefore, measurements showing a tension between the results from the 
expansion history and the results from the growth history can guide 
us to modifications of general relativity and answer the question of 
whether the acceleration comes from a new physical component or new 
physical laws. 

To see down to the deeper levels of the reality of dark energy we 
need to simultaneously fit the data for the expansion characteristics 
and the gravity theory.  Assuming one of these to be fixed can yield 
precise {\it but incorrect\/} results for both.  An illustration of 
this ``gravity's bias'' appears in Figure~1 of \cite{hl06}.  For 
mapping expansion and growth together we simulate the combination of 
the supernova (SN) distance-redshift relation and the weak gravitational 
lensing (WL) shear power spectrum (correlations between background galaxy 
shape distortions caused by light deflection by intervening mass 
structures, as a function of redshift), along with cosmic microwave 
background (CMB) data to break further degeneracies.  

\begin{figure}[!hbt]
\begin{center} 
\psfig{file=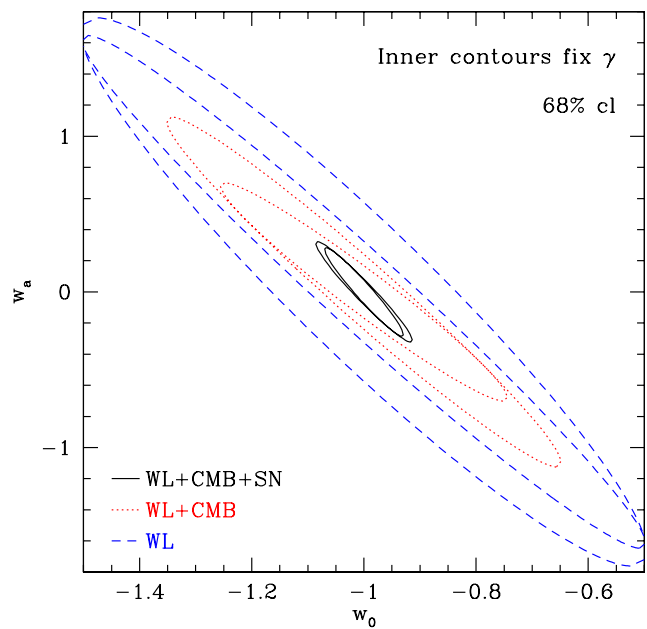,width=3.5in} 
\caption{Simultaneously fitting the cosmological expansion and 
the gravity gives robust characterization of the accelerating physics. 
The combination of weak lensing, supernovae, and cosmic microwave 
background measurements has the power to constrain the deviation from 
Einstein gravity (parameterized by $\gamma$) -- rather than assuming 
its value -- without degrading estimation of the expansion history 
parameters $w_0$, $w_a$ by more than 15-25\%. 
}
\label{fig:w0wagrav} 
\end{center} 
\end{figure}

Carrying out the analysis within the fixed gravity framework of general 
relativity can bias the result from the true answer.   
However, allowing for the possibility of deviations in 
gravity (through the gravitational growth index $\gamma$ described in 
\cite{groexp}), i.e.\ simultaneously fitting the expansion and the gravity, 
gives the correct result, with only a slight increase in the size of 
the confidence contours.  Figure~\ref{fig:w0wagrav} shows that with 
the combination of WL+SN+CMB we can indeed successfully carry out such 
a program seeking the origin of the new physics.

Furthermore, 
for the minimal modified gravity considered here, where the deviation 
from Einstein gravity is described by a single parameter, that deviation 
can be measured to 8\% precision.  So we do not want to ignore new 
gravity (hence 
biasing the results), or marginalize over it (hence learning little about 
it), but allow and fit for it.  Future work will be needed to address more 
generally the issue of simultaneously constraining the expansion 
characteristics and the gravity framework -- how wide a class of 
gravitational modifications can be described with how many parameters 
in a model independent manner?  As well, specific, well motivated 
gravity theories should be compared with simulated future data to 
explore how to achieve robust, unbiased fits and to estimate how well the 
new gravity can be recognized.  Of course not only the gravitational 
modifications enter the predictions but all the other influences on 
the growth history, 
such as neutrino mass (included in Figure~\ref{fig:w0wagrav}), running 
spectral index of the mass fluctuations, etc.  Experiments need to 
cover a wide range of spatial scales, i.e.\ high resolution and large 
area, precisely and accurately, and 
interpretation of the data must be well enough understood, to have 
confidence in any new physics.

\section{Ideas vs.\ Reality} \label{sec:ivsr}

In the previous two sections we saw there exist avenues for theoretical 
and experimental progress in revealing the new physics behind cosmic 
acceleration.  For the data, the limiting factor will be systematic 
uncertainties in the measurements, calibration, and astrophysical 
effects.  Similarly, theoretical systematics exist in the interpretation 
of the data to reach the fundamental physics.  We saw a hint of this 
with ``gravity's bias''. 

In pursuing the question of whether the acceleration is caused by a 
physical component or a new physical law, we can gain perspective by 
looking historically.  In the 18th century, discrepancies with 
Newtonian gravity were identified in the orbits of the outer planets 
of our solar system: 
was the origin a new component or a new law?  The answer of course 
turned out to be a new component -- Neptune.  In the 19th century, 
discrepancies were measured in the orbit of the inner planet Mercury. 
Here the answer was a new physical law -- Einstein gravity.  In the 
20th century, galaxy rotation curves gave unexpected results.  We are 
still waiting for the definitive resolution but a new physical 
component -- dark matter -- seems likely.  So for such a mysterious 
effect as dark energy we should not assume we know its nature ahead 
of time; we must guard against theoretical analysis that depends on 
assuming characteristics of the thing we are exploring. 

Some examples of theoretical systematic uncertainties are (see 
\cite{uzanclass} for a somewhat different approach): 

\begin{itemize} 

\item Physical fluids need not be fully described by the equation 
of state.  Perturbations in the field must exist at some level (unless 
it is a true cosmological constant), described by the sound speed 
$c_s$ and anisotropic stress $\pi_s$ (see, e.g., \cite{beandore,husound}). 
If the scalar field is canonical then $c_s=1$ and $\pi_s=0$. 

\item Gravitational effects (growth of density fluctuations, deflection 
law for light, etc.) need not be fully described by a single Newton-Poisson 
potential $\Phi$.  Two potentials, $\Phi$ and $\Psi$, enter the metric 
and can include anisotropic stress $\pi_s$, as can changes to the action, 
also leading to scale and time dependence of the gravitational coupling, 
$G(k,t)$.  (See, e.g., the review by \cite{trodden} 
and references therein.) 

\item Dark energy need not be fully dark.  Coupling of it to dark matter or 
other species could affect the matter evolution and growth, e.g.\ 
$\dot\rho_m=-3H\rho_m+\Gamma$ (see, for example, \cite{amendola}). 

\end{itemize} 

Given these possibilities, for which we do not possess a rigorous measure 
of their ``unnaturalness'', what could go wrong -- purely from fundamental 
theory systematics -- in our interpretation of various types of 
cosmological probes?  In particular, let us concentrate on geometric 
probes which do not involve issues of (nonlinear) mass growth or 
gas hydrodynamics. 

The supernova distance-redshift relation comes from the 
Friedmann-Robertson-Walker metric.  That's all.  Issues 
of dark energy sound speed, anisotropic stress, matter growth, etc.\ do 
not affect this probe. 

Weak lensing geometric probes need to separate out the mass power 
spectrum aspects of the measurements \cite{bernjain,huisteb}, which 
may be complicated by 
allowing for alterations to gravity, but may be feasible.  Otherwise 
they depend on the light deflection law, which will depend on the 
gravity theory \cite{vivi,keeton}, specifically $\Phi-\Psi$ of the 
metric potentials mentioned above.  

Baryon acoustic oscillations (BAO) as a distance probe have been analyzed 
to date within the standard cold dark matter paradigm.  The 
assumptions are that BAO in the matter power spectrum scale according 
to the standard model, to standard gravity, and blind to the dark energy 
(other than through the geometry). 
Theoretical adjustments could arise from each of these, i.e.\ the metric 
potentials $\Phi$ and $\Psi$ (see, for example, \cite{sawicki} for 
difficulties in the braneworld case), new density perturbations due 
to $c_s$ (see, e.g., \cite{dedeo}), and altered fluctuations due to 
coupling $\Gamma$ (see, e.g., \cite{amendola,mangano}).  Substantial 
and exciting 
theoretical work lies ahead to bring this method to its rich fruition. 

Table \ref{tab:theory} summarizes these putative theoretical pitfalls.

\begin{table}[!t]
  \caption{Theory systematics can impose floors on the precision 
with which cosmological probes can usefully constrain the nature of 
the acceleration physics.  Here we consider geometric probes, the 
cleanest astrophysically, and summarize the dominant and the potential 
theory systematics (see text for details). 
}
  \label{tab:theory}
  \begin{indented} \item[]
  \begin{tabular}[t]{|c|c|c|}
    \hline
    Probe & Theory Systematic (dominant) & Theory Systematic (potential) \\ 
    \hline\hline  
    SN Ia & --- & --- \\ \hline
    WL & $\Phi-\Psi$ & $c_s$, $\pi_s$, $G(k,t)$ \\ \hline 
    BAO & $\Phi$, $\Psi$, $c_s$, $\Gamma$ & 
$\pi_s$, $G(k,t)$ \\ \hline 
  \end{tabular}
  \end{indented} 
\end{table}

As one specific example, misunderstanding of the calibration required 
for the BAO 
method -- the sound horizon at decoupling -- can impose a theory 
systematics floor on precision, or bias the result.  Due to excellent CMB 
measurements, the sound horizon is rather robust to many theory 
uncertainties \cite{eiswhite} but a loophole (as identified by 
\cite{eiswhite}) lies in the predecoupling expansion history.  This 
can be upset by couplings such as in scalar-tensor theory or early 
epochs of acceleration such as in stochastic dark energy models. 
Perhaps the simplest mechanism is a nonnegligible early dark energy 
density, as predicted by dilatation symmetry solutions to the cosmological 
constant problem \cite{wett88,ratrapjep}. 
In terms of early dark energy density $\Omega_e$, \cite{doranst} found 
a shift in the sound horizon by a factor $(1-\Omega_e)^{1/2}$. 
By itself such a shift can be precisely measured through 
CMB observations, but it would not be recognized if the distance to 
the last scattering surface were shifted as well.  Here we take into 
account both effects\footnote{There is another issue, regarding the 
effect on the gravitational potentials: depending on the clustering 
properties of dark energy, the CMB may see $(\Omega_m+\Omega_e)h^2$ 
driving the potentials, while BAO, in the more recent universe, see 
$\Omega_m h^2$ (cf.\ \cite{eiswhite}).} and see what the consequences 
are if BAO are miscalibrated. 

We consider a mocker model \cite{paths} of dark energy with low redshift 
behavior of $w_0=-0.95$ and high redshift dark energy density of 
$\Omega_e=0.03$, and compare this to a constant $w$ model with the 
same low redshift behavior and negligible early dark energy density. 
(The results are similar if we use the early dark 
energy model of \cite{doranrob} instead of a mocker model.)   The CMB 
acoustic peaks are essentially indistinguishable for the two models -- 
the acoustic multipoles agree within 0.02\%.  However the unrecognized 
shift in the sound horizon causes an offset in the BAO scale that 
varies as a function of redshift, mimicking a false cosmology.  The 
uncertainty in calibration imposes a theory induced systematic of order 1\%. 

\begin{figure}[!hbt]
\begin{center} 
\psfig{file=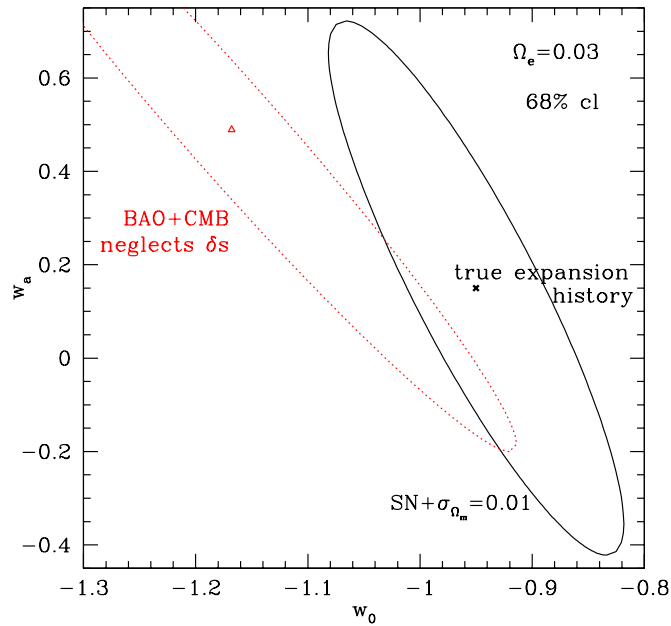,width=3.5in} 
\caption{Theory systematics can bias constraints on the expansion 
history of the universe, shown here in terms of the dark energy 
equation of state parameters $w_0=w(0)$, $w_a=-[dw/da](z=1)$. 
Miscalibration of the sound horizon $s$ due to not recognizing 
early dark energy can affect the baryon acoustic oscillation scale 
at the 1\% level, without leaving an appreciable signature in the CMB. 
The black x shows the true late universe expansion history, accurately 
mapped by supernovae.  The best fit from BAO+CMB (red triangle) 
is biased due to unrecognized miscalibration $\delta s$. 
}
\label{fig:baobias} 
\end{center} 
\end{figure}

Figure \ref{fig:baobias} illustrates the bias in cosmology parameter 
estimation caused by such sound horizon miscalibration.  The expansion 
history for the late universe, as probed by SN, does not rely on an 
early universe calibration and so is unbiased; the best fit value 
shown by the black x reproduces the distance relation for $z=0-1.7$ to 
0.2\%, or 0.004 mag.  BAO, though, suffer a 
shift in their scale and so are biased to an incorrect cosmology by 
several statistical deviations (shown by the red triangle).  
Obtaining BAO measurements more precisely than the 1\% theory 
systematics level, i.e.\ shrinking the contour in 
Figure~\ref{fig:baobias}, simply exacerbates the bias.

To some extent, the strength of BAO becomes its weakness.  BAO assumes that 
to extract the key scale from the pattern of density fluctuations, modes 
are modes are modes. 
This allows huge statistical gains by measuring huge volumes.  However, 
if modes evolve anomalously with redshift, e.g.\ because of the very 
dark energy properties we are trying to test, we will not know it. 
This contrasts with SN where its weakness turns into its strength: 
if SN properties change with redshift, we have a rich array of measurements 
to reveal this -- SN {\it are\/} distinguishable -- and so because change 
is testable we can guard against such bias.

\section{Conclusions}

Theory challenges need not be theory pitfalls or insurmountable 
obstacles to understanding the acceleration of the universe. 
We should remember Feynman's insightful quote that ``Yesterday's 
sensation is today's calibration and tomorrow's background.''  In 
the initial steps toward understanding the accelerating universe that 
we will take with the next generation of experiments we should include 
a super 
clean method like supernovae, substantially free of theory systematics 
so we obtain clear and direct interpretation in terms of fundamental 
physics.  Otherwise we run the risk of assuming the answer we are 
seeking to find.  

For further future experiments we will want more complicated 
-- rich -- dependencies so once the expansion history is well mapped 
we can probe deeper, from an established foundation, into the 
microphysical quantities like the sound 
speed and coupling and the gravitational deviations like the metric 
potentials and varying gravitational strength.  Baryon acoustic 
oscillations hold promise then 
to look at microphysical effects from both the dark matter 
and dark energy sectors.  The combination of all available probes 
will meet the theory challenges, giving a robust understanding of 
the accelerating universe.

\ack 

I thank Professor Joan Sol{\`a} and the organizing committee of IRGAC 2006 
for the invitation and hospitality.  I thank Dragan Huterer for 
collaboration leading to Figure~\ref{fig:w0wagrav}. 
This work has been supported in part by the Director, Office of Science,
Department of Energy under grant DE-AC02-05CH11231.

\end{document}